\documentclass[preprint,proceedings]{rmaa}
\usepackage{paralist}
\usepackage{psfrag,color}

\SetYear{2004}
\SetConfTitle{Proceedings of the II International Workshop on Science with GTC}
\title{Formation and Evolution of Structures in the Universe} 
\author{C. M. Guti\'errez\altaffilmark{1}, and R. Juncosa\altaffilmark{1}}
\altaffiltext{1}{Instituto de Astrof\'\i sica de Canarias, La Laguna, Tenerife, Spain.}
\shortauthor{Guti\'errez \& Juncosa}
\shorttitle{Formation and Evolution of Structures in the Universe}
\fulladdresses{
\item C.M. Guti\'errez: Instituto de Astrof\'\i sica de Canarias, 38205 La Laguna, 
Tenerife,
Spain (\email{cgc@ll.iac.es}).
}
\listofauthors{C. M. Guti\'errez}
\indexauthor{Guti\'errez, C. M.}
\abstract{

We summarize our work on structures in different stages of aggregation, and
spanning a wide range in redshift. We emphasize our  work on high resdhift
clusters.  This comprises the analysis of $\sim$7 square degrees using deep
images with  several broad-band optical filters. We plan to complement these
with  near IR observations in 4 m telescopes. Preliminary results of the
analysis in a field of 35 $\times$ 35 arcmin$^2$ indicate the detection of several
candidates to be clusters at redshifst $z> 0.5$. When all the fields are
analyzed, we will  have an adequate sample for future detailed studies with
the GTC.

}
\resumen{
Presentamos un resumen de nuestro trabajo sobre estructuras del Universo en
diferentes estados de agregaci\'on, poniendo  especial \'enfasis en el estudio
de c\'umulos de galaxias con alto corrimienro al rojo. El mismo  comprende el
an\'alisis de unos 7 grados cuadrados usando im\'agenes profundas con varios
filtros \'opticos anchos. Planeamos complementarlos con observaciones en el
infrarrojo cercano con telescopios de 4 m. Los resultados preliminares de
nuestro an\'alisis en un campo de 35x35 minutos  de arco indican la
detecci\'on de varios  candidatos a c\'umulos con corrimiento al rojo mayor
que 0.5. Cuando finalizemos el an\'alisis de todos los campos dispondremos de
una muestra adecuada para estudios detallados con GTC.

}
\addkeyword{clusters of galaxies}
\addkeyword{galaxy evolution}
\addkeyword{galaxy formation}
\addkeyword{cosmology}
\begin{document}
\maketitle

\maketitle
\section{Introduction}

In the last five years we have been conducting a project to analyze the morphology,
kinematics, and dynamics of  galaxies in different states of aggregation, from
structures similar to our Local Group (Guti\'errez et al. 2002$a$), compact groups
(Guti\'errez et al. 2002$b$), nearby clusters like Coma (Guti\'errez et al. 2004)
and low redshift clusters (Trujillo et al.  2001$b$). In these studies we have
developed new techniques for the analysis of kinematics (Prada et al.\ 1996) and
morphology (Trujillo et al.\  2001$a$) of galaxies. Here, we  summarize some of the
main results and present the current status of the project.

\section{Satellite galaxies}

The study of satellite galaxies is interesting for several reasons, one of the
most important being the discrepancy  between the observed number of such
objects and the predictions  of the standard cold dark matter (CDM) model
(Klypin et al.\ 1999, Moore et al.\ 1999); in fact, the expected  number of
satellites orbiting galaxies like the Milky Way or Andromeda is an order of
magnitude larger than that observed in the Local Group (Mateo 1998). This
could be a  strong objection against the hierarchical scenario proposed in the
standard CDM model.  To reconcile theory and observations  several mechanisms 
that suppress the formation of satellites after the re-ionization  in the
early epoch of the Universe have been proposed (e.g., Bullock, Kravtsov, \&
Weinberg 2000).  Satellite galaxies are also interesting for tracing the gravitational
potential and   for estimating the mass of the parent galaxy at distances
unreachable with other methods (Erickson, Gottesman \& Hunter 1999). The
standard models predict a decline in satellite galaxy velocity with distance
to the primary. This has been observationally explored by Zaritsky et al.\
(1997) and Prada et al.\ (2003).  

Our knowledge of satellite galaxies beyond the Local Group is still very
limited, owing to the intrinsic faintness of these objects. So detailed
studies of external systems are limited to galaxies in the nearby Universe.
The most complete compilation and study of such objects was conducted by
Zaritsky et al.\ (1997), who presented a catalogue containing 115 satellites
orbiting 69 primary isolated spiral galaxies. In Guti\'errez et al. (2002$a$)
and Guti\'errez \& Azzaro (2004) we have initiated a morphological and
photometric  analysis of an important fraction of the objects in such
catalogues. This analysis has allowed us to extend the color--color and
color--magnitude diagrams known in the Local Group. We have shown that the
$B-V$ vs.\ $M_V$ color--magnitude relation is similar to that  found in the
Fornax Cluster for galaxies with similar magnitudes. Figure~1 shows these
relations. We also have measured the internal color gradients of nine
early-type satellites and 
found   $\Delta (B-R)/\Delta (\log r)=-0.085\pm 0.027$. This study continues now with the analysis of $H_\alpha$ images to study the
star formation rate of such satellites, and especially the possible relation
with the proximity of the parent and/or other satellites.

\begin{figure}[!t]
\includegraphics[width=\columnwidth,height=5.8cm]{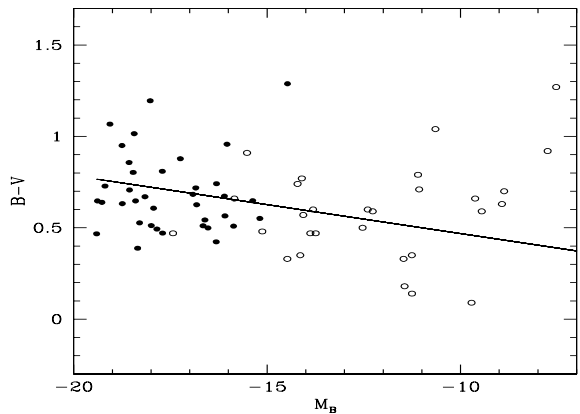}
\caption{Color--magnitude diagram of the satellite galaxies
in external galaxies ($filled\; circles$) and the dwarf galaxies of the
Local Group ($open \; circles$). The line is a least-squares fit to the
external early type satellite galaxies. }
\label{satellites}
\end{figure}

\section{Clusters of galaxies}

Clusters of galaxies are the largest gravitationally bound structures of the
Universe and  are ideal sites for testing cosmological models,  especially at
redshifts $z\ge 0.5$, where differences between competing evolutionary and
cosmological models become important (see Henry  2000 and references therein).

\subsection{Low redshift}

Since Dressler (1980), it has been known that the properties of galaxies can vary
depending  on whether they reside in dense galaxy clusters or in the field.
However, the  morphology of galaxies in clusters was mostly based on 
visual classifications. The task of quantitative analysis had left unsolved
questions such as whether the sizes of the elliptical galaxies in clusters
are different from their field counterparts, or whether the properties of spiral
galaxy disck, such as their scale-lengths, are affected by the enviroment.  In
Trujillo et al. (2001$a$) we   presented  new software for studying the
morphology of galaxies in a quantitative way. This method was applied to
the study of the Coma cluster ($z=0.02$) and A2443 ($z=0.1$) in images taken
with the  Wide Field Camera (WFC) at the Isaac Newton Telescope (INT) and
HIRAC at the Nordic Optical telewcope (NOT) respectively. In total we have
obtained the parameters of $\sim$200 galaxies up to $R\sim 17$ mag. For
the Coma Cluster  membership  was established by means of the recessional
velocity obtained from the compliation by Edwards et al.\  (2002), while
for A2443 the membership was based on the position in the color--magnitude
diagrams. 

The analysis of the Coma galaxies showed several indications of the effect
of the environment: for instance,  in Trujillo et al.\ (2002) we showed that
more concentrated elliptical galaxies tend to reside in denser regions
within the cluster, and comparing the structural properties of the Coma
cluster disk galaxies with disk galaxies in the field, we found
(Guti\'errez et al.\ 2004) that the scale lengths of the disk  galaxies in
Coma are 30\% smaller. This is illustrated in Figure~2.

\begin{figure}[!t]
\includegraphics[width=\columnwidth,height=7.8cm]{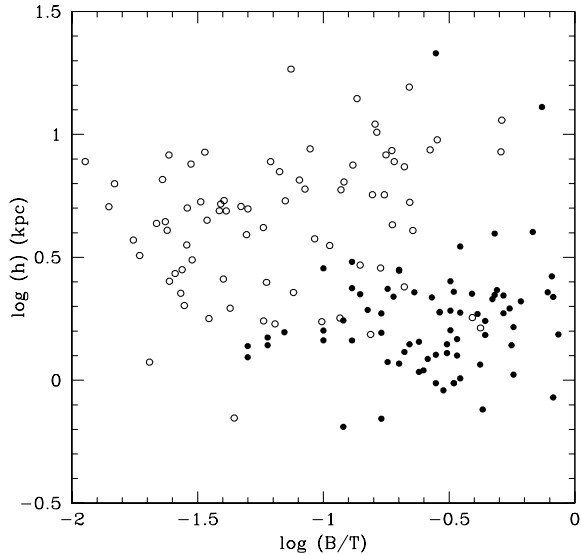}
\caption{Scale lengths of disks in the field ({\itshape open circles\/}) and in the Coma
cluster ({\itshape filled circles\/}) as a function of the bulge-to-total light
ratio $B/T$.}
\label{coma}
\end{figure}

\subsection{High redshift}

The abundance of clusters at redshift $z\sim 1$ provides complementary
information (Borgani et al.\ 2001) to that provide by the CMB anisotropies,
the Hubble diagram of  SNe 1a and the evolution  of the galaxy power
spectrum on large angular scales. This has motivated several searches for
distant clusters using a variety of techniques at different wavelengths. As
a result,  remarkable progress has been made in recent years. However, the
number of clusters discovered at high redshift is still very reduced. This
makes it difficult to conduct statistical studies spanning a wide range in 
redshift because the structures are sparse and therefore, findings require
large areas of the sky to be covered. Such large areas are not likely
to be covered in the near future with X-ray or Sunyaev--Zel'dovich effect searches.
The alternative that we are following is the use of wide field imaging in the
optical and near IR to  detect two-dimensional density enhancements
complemented with the determination of photometric redshifts. 

For optical data we use the released data of the NOAO
($http://www.noao.edu/noao/noaodeep/$) and DLS ($http://dls.bell-labs.com/$)
deep surveys. With these two public surveys we have $\sim 7$ square degrees in
several broad band optical filters (a few more degrees are expected to be
released soon). For instance the limiting  magnitude in the $R$ filter for the
NOAO data is $\sim$ 25 mag. Previous analysis in the southern survey EIS
(Benoist et al. 2002) show that the density of clusters at $z\ge 0.6$ is $\ge
10$ per square degree, so  a survey such as  the one we are analyzing 
will provide a sample representative for statistical analysis,
 and for further
spectroscopy with  GTC.

Using the Sextractor software  (Bertin \& Arnouts 1996), we have built catalogues of objects  
comprising more than 10$^5$ objects
per square degree  (mostly galaxies). We have applied several methods based
on two-dimensional density enhancements (e.g., Ramella et al. 2001) complemented
with   color and magnitude information (Gladders \& Yee 2000). We also
have applied the method of photometric redshifts using several codes
(Fern\'andez-Soto et al. 1999; Bolzonella et al. 2000). The methods have
been tested by comparing with the known spectroscopic redshifts of objects in the
field. This comparison is shown in Figure~3. We found a reasonably good
agreement between spectroscopic and photometric redshifts. The rms is
$\Delta z\sim 0.1$, although  there is a small systematic effect that
tends to make the photometric redshift slightly smaller than the real one.
We are investigating the reasons for this.

\begin{figure}[!t]
\includegraphics[width=\columnwidth,height=6.2cm]{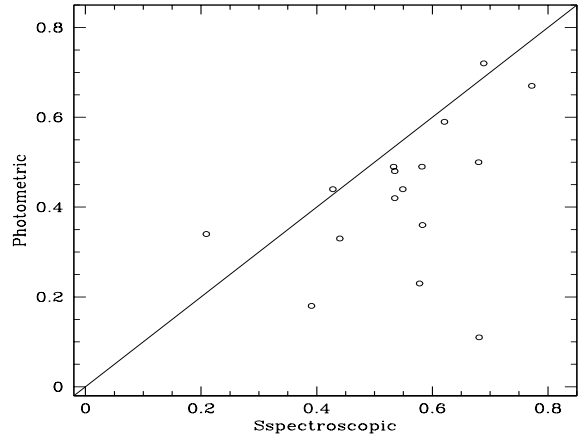}
\caption{Comparison between spectroscopic and photometric redshifts of
galaxies in one of the DLS fields.}
\label{coma}
\end{figure}

\begin{figure}[!t]
\includegraphics[width=\columnwidth,height=6.2cm]{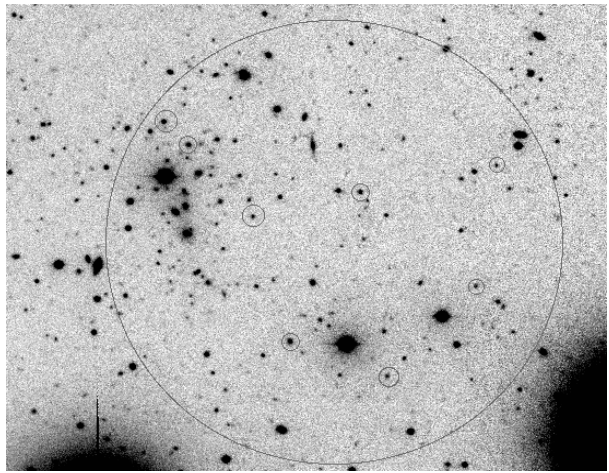}
\caption{Image in the $z^\prime$ filter  centered on the position in which we
identified a cluster at $z\sim 0.9$ in a subfield of the DLS survey.  The large circle has a diameter of 3
arcmin and encloses the region where the cluster members are. Galaxies identified as member of such
cluster are enclosed in the small circles.}
\label{coma}
\end{figure}

So far, we have analyzed one of the subfields of the DLS survey (see Juncosa
et al. 2004, this volume) and have identified several  good candidate 
clusters at $z\ge 0.5$. Figure~4 shows one of the candidates. However, the
typical uncertainties in redshift at $z<1$ are $\Delta z\sim 0.1$ which
makes the detection of clusters difficult and produces some spurious
identifications. This could be largely improved by including observations in the 
near IR. We therefore plan to complement these optical surveys with a wide
near IR survey taking advantage of the capabilities and in particular the wide
field of view of $\Omega-2000$ at the 3.5 m telescope at Calar Alto. This  is
one of the very few existing instruments  that permits deep surveys (limiting
magnitude $K^\prime \sim 19.5)$ of a few square degrees in the near IR
in a reasonable exposure time.  Follow-up of the structures detected  will be
carried out with spectroscopy of the member galaxies  of these clusters  with
the Spanish GTC  and its first-light instrument OSIRIS
($http://www.iac.es/proyect/OSIRIS$). This will allow a deep study of physical
properties such as their  ages, metallicities and   evolution according to 
redshift.


\begin{thebibliography}{}
\bibitem[Benoist et al. (2002)]{ben02} Benoist, C. et al. 2002, A\&A, 394, 1
\bibitem[Bertin \& Arnouts (1996)]{ber96}Bertin, E., \& Arnouts, S. 1996 A\&AS, 117, 393    
\bibitem[Bolzonella et al. (2000)]{bol00}Bolzonella, M., Miralles, J. M., \& Pell\'o, R. 2000, A\&A, 363, 476   
\bibitem[Borgani et al. (2001)]{bor01}Borgani, S. et al. 2001, ApJ, 561, 13   
\bibitem[Bullock et al. (2000)]{bul00}Bullock, J. S., Kravtsov, A. V., Weinberg, D. H. 2000, ApJ, 539, 517  
\bibitem[Dressler (1980]{dre80} Dressler, A. 1980, ApJSS, 42, 565   
\bibitem[Edwards et al. (2002)]{edw02}Edwards, S. A., Colless, M., Bridges, T. J., Carter, D., 
             Mobasher, B., \& Poggianti, B. M. 2002, ApJ, 567, 178  
\bibitem[Erickson et al. (1999)]{eri99} Erickson, L. K., Gottesman, S. T., \& Hunter, Jr., J. H. 1999, ApJ,
515, 153  
\bibitem[Fern\'andez-Soto et al. (1999)]{fer99}Fern\'andez-Soto, A., Lanzetta, K. M., \& Yahil, A. 1999, ApJ, 513, 34  
\bibitem[Gladders \& Yee (2000){gla00}Gladders, M. D., \& Yee, H. K. C. 2000, ApJ, 120, 2148   
\bibitem[Guti\'errez \& Azzaro (2004)]{gut04}Guti\'errez, C. M., \& Azzaro, M. 2004,
ApJ (accepted), astro-ph/0404587
\bibitem[Guti\'errez et al. (2002$a$){gut02a}Guti\'errez, C. M., Azzaro, M., \& Prada, F. 2002$a$, ApJS, 141, 61  
\bibitem[Guti\'errez et al. (2002$b$]{gut02b}Guti\'errez, C. M., L\'opez-Corredoira, M., Prada, F., \& Eliche, M. C.
2002$b$ ApJ, 579, 592  
\bibitem[Guti\'errez et al. (2004]{gutn04}Guti\'errez, C. M., Trujillo, I., Aguerri, J. A. L., Graham, A.
W., \& Caon, N. 2004, ApJ, 602, 664  
\bibitem[Henry (2000)]{hen00}Henry, J. P. 2000, ApJ, 534, 565   
\bibitem[Juncosa \& Guti\'errez (2004]{jun04}Juncosa, R., Guti\'errez, C. M., \& Fern\'andez-Soto, A. 2004 (this
volume)   
\bibitem[Klypin et al. (1999)]{kly99}Klypin, A., Kravtsov, A. V., Valenzuela, O., \& Prada, F. 1999, ApJ, 522,
82  
\bibitem[Mateo (1998)]{mat98}Mateo, M. L. 1998, ARA\&A, 36 435  
\bibitem[Moore et al. (1999)]{moo99}Moore, B., Ghigna, S., Governato, F., Lake, G., Quinn, T., Stadel, J., \& Tozzi, P. 1999, 
ApJ, 524, L19  
\bibitem[Prada et al. (2003)]{pra99}Prada, F. et al. 2003, ApJ, 598, 260  
\bibitem[Prada et al. (1996)]{pra96}Prada, F., Guti\'errez, C. M., Peletier, R. F., \& McKeith, C. D.  1996,
ApJ, 463, L9   
\bibitem[Ramella et al. (2001)]{ram01}Ramella, M., Boschin, W., Fadda, D., \& Nonino, M. 2001, A\&A, 368, 776   
\bibitem[Trujillo et al. (2001$a$)]{tru01a}Trujillo, I., Aguerri, J. A. L., Cepa, J., \& Guti\'errez, C. M., 
2001$a$, MNRAS, 321, 269  
\bibitem[Trujillo et al. (2001$b$)]{tru01b}Trujillo, I., Aguerri, J. A. L., Guti\'errez, C. M., \& Cepa, J.
2001$b$, MNRAS, AJ, 122, 38    
\bibitem[Trujillo et al  (2002)]{tru02}Trujillo, I., Aguerri, J. A. L., Guti\'errez, C. M., Caon, N., \& Cepa, J.
2002, ApJ, 573, L9  
\bibitem[Zaritsky et al. (1997)]{zar97}Zaritsky, D., Smith, R., Frenk, C., \& White, S. D. M. 1997, ApJ, 478, 39
  
\end{thebibliography}
\end{document}